\documentstyle[preprint,aps,epsfig]{revtex}
\tightenlines
\begin{document}
\title{Infrared behavior of the running coupling constant
and bound states in QCD}
\author{M. Baldicchi, G. M. Prosperi}
\address{Dipartimento di Fisica dell'Universit\`{a} di Milano \\
and I.N.F.N., Sezione di Milano, Italy \\}
\date{\today}
\maketitle
\begin{abstract}
The perturbative expression of the running strong coupling constant
$ \alpha_{\rm s}( Q^{2} ) $ has an unphysical singularity for
$ Q^{2} = \Lambda^{2}_{\rm QCD} $. Various modification have
been proposed for the infrared region. The effect of some of such
proposals on the quark-antiquark spectrum is tested on a
Bethe-Salpeter (second order) formalism which was successfully
applied in previous papers
to an overall evaluation of the spectrum in the light-light,
light-heavy and heavy-heavy sectors (the only
serious discrepancy with data
being for the light pseudoscalar meson masses).
In this paper only the $ {\rm c} \bar{\rm c} $,
$ {\rm b} \bar{\rm b} $ and $ {\rm q} \bar{\rm q} $
(q = u or d) cases are
considered and fine structure is neglected. It is found that in the
$ {\rm b} \bar{\rm b} $ and $ {\rm c} \bar{\rm c} $ cases the results
are little sensitive to the specific choice.
In the light-light case the Dokshitzer $ et \; al. $ prescription
is again
essentially equivalent to the truncation prescription used in the
previous calculation and it is consistent with the same
$ a \; priori $ fixing of the quark light
masses on the typical current
values $ m_{\rm u} = m_{\rm d} = 10 \; {\rm MeV} $
(only the pion mass
resulting completely out of scale of about 500 MeV).
With the Shirkov-Solovtsov prescription, on the contrary, a
reasonable agreement with the data is obtained only at the price of
using a phenomenological momentum dependent effective mass
for the quark.
The use of such an effective mass should amount to a correction of
the free quark propagator.
It is remarkable that this has also the effect of
bringing the pion mass in the correct range.
\end{abstract}
\setcounter{equation}{0}
\newpage
\section{INTRODUCTION}
In perturbation theory the running coupling constant in
QCD is usually written up to one loop as
\begin{equation}
  \alpha_{\rm s} ( Q^{2} ) = \frac{ 4 \pi }{ \beta_{0}
  \ln{ ( Q^{2} / \Lambda^{2} ) } }
\label{runcst}
\end{equation}
or also up to two loops
\begin{equation}
  \alpha_{\rm s} ( Q^{2} ) = \frac{ 4 \pi }{ \beta_{0}
  \ln{ ( Q^{2} / \Lambda^{2} ) } }
  \left[ 1 + \frac{ 2 \beta_{1} }{ \beta_{0}^{\, 2} }
  \frac{ \ln ( \ln{ ( Q^{2} / \Lambda^{2} ) ) } }
  { \ln{ ( Q^{2} / \Lambda^{2} ) } }
  \right] ,
\label{runcst2}
\end{equation}
$ Q $ being the relevant energy scale,
$ \beta_{0} = 11 - \frac{2}{3}  n_{\rm f} $,
$ \beta_{1} = 51 - \frac{19}{3} n_{\rm f} $ and
$ n_{\rm f} $ the number of flavors
with masses smaller than $ Q $.

Such expressions have been largely tested in the large
$ Q $ processes and are normally used to relate data
obtained at different
$ Q $ using the appropriate number of
``active'' flavors $ n_{\rm f} $ and different values of
$ \Lambda $ in the ranges between the various quark thresholds.

Both expressions become singular and completely inadequate as
$ Q^{2} $ approaches $ \Lambda^{2} $.
Therefore they must be somewhat modified in the
infrared region.

Various proposals have been done in this direction.
The most naive assumption consists in cutting the
curve (\ref{runcst}) at a certain maximum value
$ \alpha_{\rm s}(0) = \bar{\alpha}_{\rm s} $
to be treated as a mere phenomenological parameter
(truncation prescription).
Alternatively,
on the basis of general analyticity arguments,
Shirkov and Solovtsov \cite{shirkov} replace
(\ref{runcst}) with
\begin{equation}
  \alpha_{\rm s} ( Q^{2} ) = \frac{ 4 \pi }{
   \beta_{0} } \left(
  \frac{1}{ \ln{ ( Q^{2} / \Lambda^{2} ) } } +
  \frac{ \Lambda^{2} }{ \Lambda^{2} - Q^{2} } \right).
\label{runshk}
\end{equation}
This remains regular for $ Q^{2} = \Lambda^{2} $ and
has a finite $ \Lambda $ independent limit
$ \alpha_{\rm s}(0) = 4 \pi / \beta_{0} $,
for $ Q^{2} \rightarrow 0 $.
Finally, inspired also by phenomenological reasons,
Dokshitzer $ et \; al. $ \cite{lucenti} write
\begin{equation}
  \alpha_{\rm s} ( Q^{2} ) =
  \frac{ \sin ( \pi {\cal P} ) }{ \pi {\cal P} } \;
  \alpha^{0}_{\rm s} ( Q^{2} ),
\label{runlcnt}
\end{equation}
where $ \alpha^{0}_{\rm s} ( Q^{2} ) $ is the perturbative
running coupling constant as given by Eq. (\ref{runcst})
and
$ {\cal P} = d/d ( \ln ( Q^{2} / \Lambda^{2} ) ) $
is a derivative acting on
$ \alpha^{0}_{\rm s} ( Q^{2} ) $.
The various curves are reported in Fig. \ref{fig1}.

The above modified expressions have been applied to study
various effects in which infrared behavior
turns out to be important.
Electron-positron annihilation into hadrons, $ \tau $-lepton
decay, lepton-hadron deep inelastic scattering, jet shapes,
pion form-factors etc. are of this type.

In the quark-antiquark bound state problem the variable $ Q^{2} $
can be identified with the squared momentum transfer
$ {\bf Q}^{2} = ( {\bf k} - {\bf k}^{\prime} )^{2} $ and formally
the use of a running coupling constant amounts to include higher
order terms in the perturbative part of the potential or the
Bethe-Salpeter kernel. In this case all values of $ {\bf Q}^{2} $
are involved and an infrared regularization becomes essential.
Furthermore $ \langle {\bf Q}^{2} \rangle $
ranges typically between
$ ( 1 \, {\rm GeV} )^{2} $ and $ ( 0.1 \, {\rm GeV} )^{2} $
for different quark masses and internal excitations
and values of $ {\bf Q}^{2} $ smaller than
$ \Lambda^{2} $ can be important.
The specific infrared behavior is therefore expected
to affect the spectrum and other properties of mesons.

The purpose of this paper is to test such kind of
effects in a particular
formalism we have developed and used in previous papers.

\section{FORMALISM}
In reference \cite{quadratico} we have obtained a good reproduction
of the entire meson spectrum in terms of only four adjustable
parameters, by solving numerically the eigenvalue equation for
the squared mass operator
\begin{equation}
  {\rm M}^{2} = {\rm M}_{0}^{2} + {\rm U}
\label{eqM2}
\end{equation}
or the mass operator
\begin{equation}
  {\rm M} = {\rm M}_{0} + {\rm V},
\label{eqM1}
\end{equation}
where $ {\rm M}_{0} = w_{1} + w_{2} =
\sqrt{ m_{1}^{2} + {\bf k}^{2} } +
\sqrt{ m_{2}^{2} + {\bf k}^{2} } $ is the kinetic term and
U and V are complicated momentum dependent potentials.
Up to the first order in the running coupling constant
$ \alpha_{\rm s} ( {\bf Q}^{2} ) $ and in terms of
the string tension
$ \sigma $, the ``quadratic potential'' U is given by
\begin{eqnarray}
&&         \! \! \! \!
  \langle {\bf k} \vert U \vert {\bf k}^\prime \rangle =
  \sqrt{ \frac{ ( w_{1} +w_{2} ) ( w_{1}^{\prime}
  + w_{2}^{\prime} ) }{
  w_{1} w_{2} w_{1}^{\prime} w_{2}^{\prime} } }
  \bigg\{ {4\over 3} 
\frac{ \alpha_{\rm s} ( {\bf Q}^{2} ) }{ \pi^2 }
  \bigg[ - {1\over {\bf Q}^2}
  \bigg( q_{10} q_{20} + {\bf q}^2 -
  { ( {\bf Q} \cdot {\bf q})^2 \over {\bf Q}^2 } \bigg) +
\nonumber \\
&&  + \frac{i}{ 2 {\bf Q}^{2} }
  {\bf k} \times {\bf k}^{\prime} \cdot (
  \mbox{\boldmath $ \sigma $}_{1} +
  \mbox{\boldmath $ \sigma $}_{2} )
  + \frac{1}{ 2 {\bf Q}^2 } [ q_{20}
  ( \mbox{\boldmath $ \alpha $}_{1}
  \cdot {\bf Q} ) - q_{10} (
  \mbox{\boldmath $ \alpha $}_{2}
  \cdot {\bf Q}) ] +
\nonumber \\
&&  + \frac{1}{6}
  \mbox{\boldmath $ \sigma $}_{1}
  \cdot
  \mbox{\boldmath $ \sigma $}_{2}
  + \frac{1}{4} \left( \frac{1}{3}
  \mbox{\boldmath $ \sigma $}_{1}
  \cdot
  \mbox{\boldmath $ \sigma $}_{2} -
  \frac{ ( {\bf Q} \cdot
  \mbox{\boldmath $ \sigma $}_{1} )
  ( {\bf Q} \cdot
  \mbox{\boldmath $ \sigma $}_{2} ) }{
  {\bf Q}^2 } \right)
  + \frac{1}{ 4 {\bf Q}^2 } (
  \mbox{\boldmath $ \alpha $}_{1}
  \cdot {\bf Q} ) (
  \mbox{\boldmath $ \alpha $}_{2}
  \cdot {\bf Q} ) \bigg] +
\nonumber \\
&&  + \int \!
  \frac{ d^{3} {\bf r} }{ ( 2 \pi)^3 } \,
  e^{ i {\bf Q} \cdot {\bf r} }
  J^{\rm inst}( {\bf r} , {\bf q} , q_{10} , q_{20})  \bigg\},
\label{eq:upot}
\end{eqnarray}
with
\begin{eqnarray}
&& \! \! \! \! \! \!
  J^{\rm inst} ( {\bf r} , {\bf q} , q_{10} , q_{20} ) =
  \frac{ \sigma r }{ q_{10} + q_{20} }
  \bigg[ q_{20}^{2} \sqrt{ q_{10}^{2} - {\bf q}_{\rm T}^{2} }
  + q_{10}^{2} \sqrt{ q_{20}
  - {\bf q}_{\rm T}^{2} } +
\nonumber  \\
&& \! \! \! \! \! \!
  + \frac{ q_{10}^{2} q_{20}^{2} }{ \vert {\bf q}_{\rm T} \vert }
  \big( \arcsin \frac{ \vert {\bf q}_{\rm T} \vert }{ q_{10} }
  + \arcsin \frac{ \vert {\bf q}_{\rm T} \vert }{ q_{20} }
  \bigg) \bigg]
  - \frac{ \sigma }{r} \bigg[ \frac{ q_{20} }{
  \sqrt{ q_{10}^{2} - {\bf q}^{2}_{\rm T} } }
  ( {\bf r} \times {\bf q} \cdot
  \mbox{\boldmath $ \sigma $}_{1} +
\nonumber  \\
&& \! \! \! \! \! \!
  + i q_{10} ( {\bf r} \cdot
  \mbox{\boldmath $ \alpha $}_{1} ) )
  + \frac{ q_{10} }{ \sqrt{ q_{20}^{2} - {\bf q}^{2}_{\rm T} } }
  ( {\bf r} \times {\bf q} \cdot
  \mbox{\boldmath $ \sigma $}_{2}
  - i q_{20} ( {\bf r} \cdot
  \mbox{\boldmath $ \alpha $}_{2} ) ) \bigg]
\label{eq:uconf1}
\end{eqnarray}
and
$
  \alpha_{j}^{k} = \gamma_{j}^{0} \gamma_{j}^{k}, \;
  \sigma_{j}^{k} = \frac{i}{4} \;
  \varepsilon^{knm} [ \gamma^{n}_{j} , \gamma^{m}_{j} ], \;
  q_{j0} = ( w_{j} + w_{j}^{\prime} )/2, \; j = 1,2
$.
$
  {\bf Q} = {\bf k} - {\bf k}^{\prime}
$, \\
$
  {\bf q} = ( {\bf k} + {\bf k}^{\prime} )/2, \;
  q^{h}_{\rm T} = ( \delta^{hk} -
  \hat{r}^{h} \hat{r}^{k} )q^{k}
$,
$
  \hat{\bf r} = {\bf r}/r
$.

The above expression was obtained by reducting of a Bethe-Salpeter
like equation which was obtained in reference \cite{bmp} from
first principle QCD under the only assumption that the logarithm
of the Wilson loop correlator $ W $ could be written as the sum of
its perturbative expression and an area term
\begin{equation}
  i \ln W = i ( \ln W )_{\rm pert} + \sigma {\rm S}
\end{equation}
(advantage was taken in the derivation of an appropriate
Feynman-Schwinger representation of the quark propagator in an
external field).

An expression for
$ \langle {\bf k} \vert V \vert {\bf k}^\prime \rangle $
can be obtained by a direct comparison of
Eq. (\ref{eqM1}) with Eq. (\ref{eqM2}).
Neglecting terms in $ V^{2} $ (what is consistent with the other
approximations), this amounts simply to change the kinematical
factor in front of the right-hand side of Eq. (\ref{eq:upot}).
Properly one should divide
$ \langle {\bf k} \vert U \vert {\bf k}^\prime \rangle $
by $ w_{1} + w_{2} +  w_{1}^{\prime} + w_{2}^{\prime} $,
practically the simpler replacement
$
  \sqrt{ \frac{ ( w_{1} + w_{2} )
  ( w_{1}^{\prime} + w_{2}^{\prime} ) }{ w_{1} w_{2}
  w_{1}^{\prime} w_{2}^{\prime} } }
\to \frac{1}{ 2\sqrt{ w_{1} w_{2} w_{1}^{\prime} w_{2}^{\prime} } }
$
is essentially equivalent.

The interest of the more conventional Eq. (\ref{eqM1}) is
that it makes more immediate a comparison with
ordinary potential approaches
and the consideration of the non-relativistic limit.
In particular $ V $ coincides with the Cornell potential
$ \langle {\bf k} \vert V \vert {\bf k}^\prime \rangle =
\langle {\bf k} \vert ( - \frac{4}{3} \frac{
\alpha_{\rm s} }{r} + \sigma r )
\vert {\bf k}^\prime \rangle $
in the static limit 
and with the potential obtained in \cite{barch} when
the first relativistic corrections are included.
In this paper, however,
we shall refer only to Eq. (\ref{eqM2}), as more directly
related to the original B-S equation.

The method used in \cite{quadratico} consists in solving first the
eigenvalue equation for M in the static limit of
$ V $ by the Rayleigh-Ritz method (using an harmonic oscillator
basis); then in evaluating
the quantity $ \langle {\rm M}^{2} \rangle $
(or $ \langle {\rm M} \rangle $)
for the resulting zero order eigenfunctions.

Actually in \cite{quadratico}
we have neglected the complicated fine
structure spin dependent terms occurring in (\ref{eq:upot})
and taken into account only the hyperfine term in
$ \frac{1}{6} \, \mbox{\boldmath $ \sigma $}_{1} \!
 \cdot \mbox{\boldmath $ \sigma $}_{2} $.
We have used the expression (\ref{runcst}) for the running
coupling constant with $ n_{\rm f} = 4 $ and
$ \Lambda = 0.2 \; {\rm GeV} $ frozen
at $ \bar{\alpha}_{\rm s} = 0.35 $; we have also taken
$ \sigma = 0.2 \; {\rm GeV}^{2} $, $ m_{\rm u} = m_{\rm d} = 10 \;
{\rm MeV} $ \footnote{Notice that the results are very little
sensitive to the precise values of $ m_{\rm u} $ and $ m_{\rm d} $
if these are small.}, $ m_{\rm c} = 1.394 \; {\rm GeV} $,
$ m_{\rm b} = 4.763 \; {\rm GeV} $. The quantities $ \Lambda $
and $ m_{\rm u} = m_{\rm d} $ were fixed $ a \; priori $ from high
energy data; $ \bar{\alpha}_{\rm s} $, $ \sigma $, $ m_{\rm c} $ and
$ m_{\rm b} $ were adjusted on the ground $ c \bar{c} $ and
$ b \bar{b} $ states, the $ c \bar{c} $ hyperfine splitting and
the Regge trajectory slope.

In this paper, to test the sensitivity of the results to the
infrared behavior of $ \alpha_{\rm s} ( {\bf Q}^{2} ) $, we have
performed the same calculation in the $ b \bar{b} $, $ c \bar{c} $
and $ q \bar{q} $ ($ q = u $ or {\sl d})
case using Eqs. (\ref{runshk}) and (\ref{runlcnt}),
with an appropriate redefinition of the adjustable parameters.

\section{RESULTS}
In tables \ref{tablbb}, \ref{tablcc}, and \ref{tabluu} we give the
$ b \bar{b} $, $ c \bar{c} $, and $ q \bar{q} $ quarkonium masses
respectively obtained for the different running coupling constant
prescriptions.
In column (a) we report the results obtained in ref. \cite{quadratico}
for the truncated $ \alpha_{\rm s} ( {\bf Q}^{2} ) $,
in column (b) those obtained by means of Eq. (\ref{runshk})
proposed by Shirkov-Solovtsov and
in column (c) those obtained by means of the
$ \alpha_{\rm s} ( {\bf Q}^{2} ) $ of Eq. (\ref{runlcnt})
proposed by Dokshitzer $ et \; al. $

In columns (b) and (c) we used the same values
$ n_{\rm f} = 4 $, $ \Lambda = 0.2 \; {\rm GeV} $ and
$ m_{\rm u} = m_{\rm d} = 10 \; {\rm MeV} $ as in \cite{quadratico},
but we have slightly redefined the adjustable parameters taking
$ \sigma = 0.18 \; {\rm GeV}^{2} $ in both cases and
$  m_{\rm c} = 1.545 \; {\rm GeV} $ and
$  m_{\rm b} = 4.898 \; {\rm GeV} $ for prescription (\ref{runshk})
(column (b)), $  m_{\rm c} = 1.383 \; {\rm GeV} $ and
$  m_{\rm b} = 4.7605 \; {\rm GeV} $ for prescription
(\ref{runlcnt}) (column (c)).

Notice that, in spite of the reduced number of adjustable
parameters, the spectra of bottonium and charmonium are not
essentially modified by the new choice for
$ \alpha_{\rm s} ( {\bf Q}^{2} ) $, with perhaps the exceptions of
the highest $ c \bar{c} $ states that are lower in
the Dokshitzer  $ et \; al. $ case.
This indicate little sensitivity of such spectra
to the infrared behavior of $ \alpha_{\rm s} ( {\bf Q}^{2} ) $.

The situation is completely different for the light-light spectrum
of table \ref{tabluu}. While in front of the experimental and
theoretical uncertainties columns (a) and (c) can be considered
not really distinguishable, the values reported in column (b) are
definitely systematically too low (in particular
$ \langle {\rm M}^{2} \rangle < 0  $ for $ \pi $-meson).

Notice however that in the above reported calculations we have used
for $  m_{\rm u} $ and $  m_{\rm d} $ the current mass value of
$ 10 \; {\rm MeV} $. This amounts to assume the difference
between the current and the constituent masses to be essentially
related to kinematical relativistic correction
(cf. \cite{quadratico}) or, what is the same,
that the free quark propagator is a good approximation
for the complete one in the B-S equation.
The inability of the formalism to reproduce a reasonable
value for the $ \pi $ mass and all experience gained by the chiral
symmetry problematic (see in particular \cite{chiral} and reference
herein) suggests that this should not be the case for the light-light
systems.

For this reason we have repeated the calculation for choice
(\ref{runshk}) with various constituent values
for the light quark masses. Two sets of results are reported in
columns (d) and (e) of table \ref{shrkuu}.
Notice that for $ m_{\rm u} = m_{\rm d} = 0.30 \; {\rm GeV} $
the situation is again very
similar to those of column (a), table \ref{tabluu}.
Notice also however
that as $ m_{\rm u,d} $ increases the bound state masses
uniformly increase and that for low value of $ m_{\rm u,d} $ the
lowest bound state masses can be made to agree fairly well with
the data, for high value the same occurs for higher states.
This could suggest the use of a kind of running constituent mass.

In column (h) the results are reported for the squared effective
mass
\begin{equation}
  m_{\rm eff}^{2} = 0.11 \, k - 0.025 \, k^{2} + 0.265 \, k^{4}
\label{massrun1}
\end{equation}
$ k $ denoting the quark momentum in the center of mass frame.
In Eq. (\ref{massrun1}) the coefficients are chosen in order
to obtain $ m_{\rm eff} = 0.22, \, 0.28, \, 0.35 \; {\rm GeV} $ for
$ k^{2} = 0.26, \, 0.41, \, 0.58 \; {\rm GeV}^{2} $
approximately corresponding to the $ \langle {k}^{2} \rangle $
values for the 1S, 2D and 1G states respectively.
As it can be seen the
agreement with the data is much improved in this way and finally
even a reasonable value for the $ \pi $ mass is obtained.

Notice that the use of an effective running mass is in agreement
with the general perspectives of combining Dyson-Schwinger
equation with B-S equation.
Obviously a fine tuning of the coefficients in Eq. (\ref{massrun1})
to further improve the results would be meaningless in the
present context, due even to the approximation used
({\sl e.g.} the triplet-singlet splitting
has been essentially evaluated perturbalively).

\section{CONCLUSIONS AND DISCUSSION}

In conclusion the heavy quarkonium spectrum does not seem to be very
sensitive to the specific infrared behavior of the running constant
as it could be expected on general ground.

\begin{figure}[htbp!]
  \begin{center}
    \leavevmode
    \setlength{\unitlength}{1.0mm}
    \begin{picture}(170,150)
      \put(0,0){\mbox{\epsfig{file=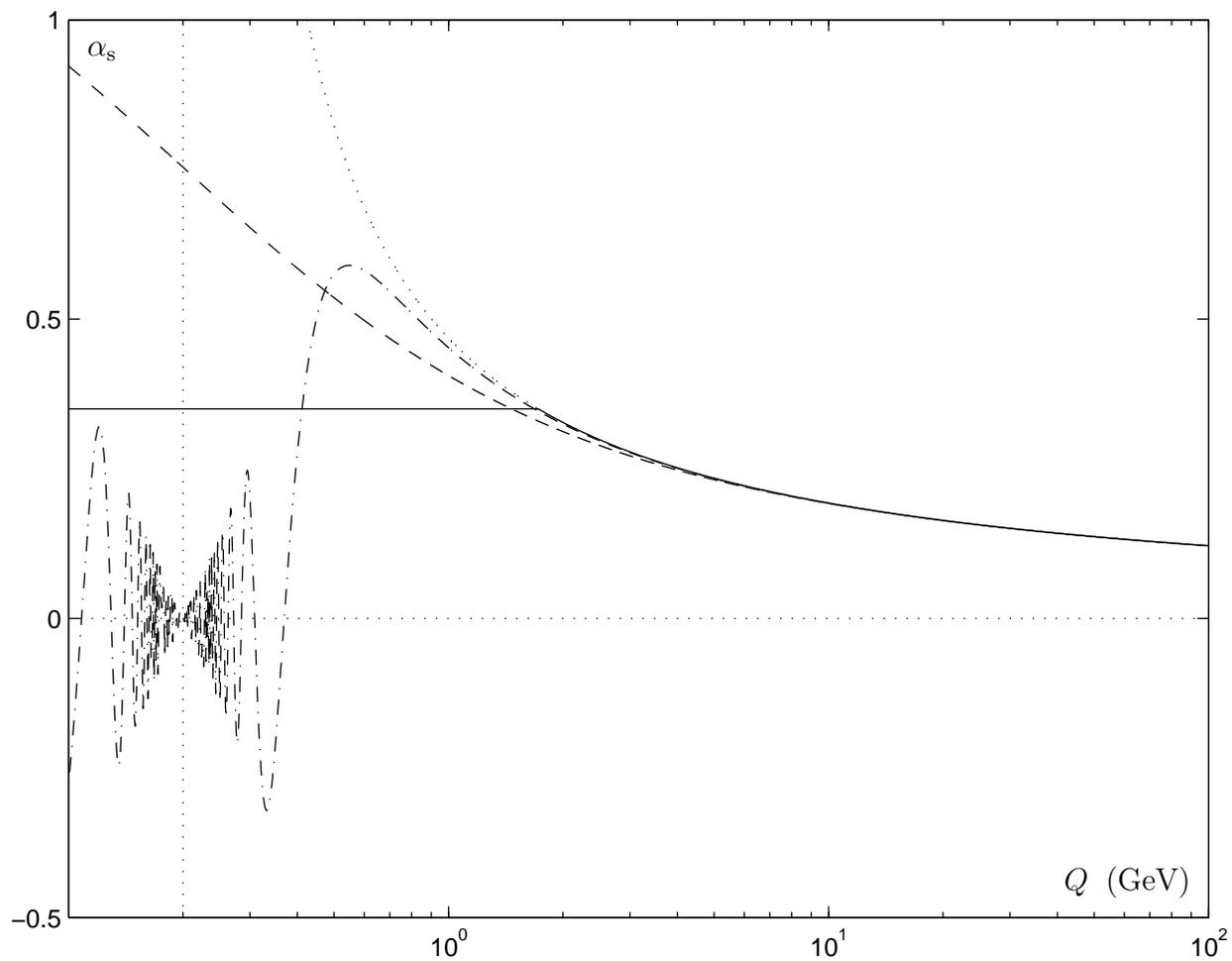,height=13cm}}}
      \put(9,122){ $ \alpha_{\rm s} $ }
      \put(140,10){ $ Q $ }
      \put(147,10){(GeV)}
    \end{picture}
  \end{center}
\vspace{7mm}
\caption{Running coupling constant $ \alpha_{\rm s} (Q) $
on logarithmic scale.
Truncation prescription (full line),
Shirkov-Solovtsov prescription (dashed line),
Dokshitzer $ et \; al. $ prescription (dot-dashed line).}
\label{fig1}
\end{figure}

For the light quarkonium Dokshitzer $ et \; al. $ prescription
(\ref{runlcnt}) does not essentially change the results in comparison
with the truncation assumption adopted in \cite{quadratico},
in spite of the very different appearance of the corresponding curve
in Fig. \ref{fig1}

On the contrary the situation changes drastically for
prescription (\ref{runshk}).

Actually such prescription would be definitely ruled out if we insisted
in using a current mass for the light quarks. If however we give
to $ m_{\rm u,d} $ a constituent value the results are again similar
to those obtained with the truncated
$ \alpha_{\rm s} ( {\bf Q}^{2} ) $ and can be strongly improved
if we use the running effective mass (\ref{massrun1}).

To understand better the meaning of Eq. (\ref{massrun1}),
notice that, in the context of the second
order formalism developed
in ref. \cite{bmp}, the free quark propagator $ H_{2}^{(0)}(p) $
occurring in the BS equation is $ i/(p^{2} - m^{2}) $,
where we have to set
$ p = ( M_{\rm B}/2 \pm k_{0}, \pm {\bf k} ) $ (the upper and
lower signs referring to the quark and the antiquark respectively)
in the C.M. frame .
However if, consistently with the other approximations made,
we neglect the spin dependent terms,
the full quark propagator $ H_{2}(p) $ can be written as
$ i/( p^{2} - m^{2} + \Gamma(p) ) $.
Then, recalling that
the instantaneous approximation consists in setting $ k_{0} = 0 $
in the BS kernel, in the same order of ideas we can replace the slowly
varying expression $ \Gamma( p_{0}, \vert {\bf p} \vert ) $ by
$ \Gamma( M_{\rm B}/2, \vert {\bf k} \vert ) $.
Eventually we obtain the operator $ M^{2} $ as given by
Eq. (\ref{eqM2}) but with $ m_{1} = m_{2} $ replaced by
\begin{equation}
  m_{\rm eff}^{2} ( \vert {\bf k} \vert ) = m^{2} - \Gamma( M_{\rm B}/2,
  \vert {\bf k} \vert ).
\label{massrun2}
\end{equation}
Eq. (\ref{massrun1}) corresponds to a parametrization of right hand
side of Eq. (\ref{massrun2}). Obviously in principle
$ \Gamma( p ) $ should be obtained by actually
solving the DS equation for
$ H_{2}(p) $ (ref. [3b,4b,4c]), but this is a complicated task that
we reserve to a forthcoming paper.

\acknowledgments

We gratefully acknowledge a discussion with Prof. Shirkov who
recalled our attention on the interest of the problem.
\begin{table}
\centering
\caption{$ b \bar{b} $.
$ n_{f} = 4, \; \Lambda = 0.2 $ GeV.
(a) $ m_{\rm b} = 4.763 \; {\rm GeV} $,
$ \sigma = 0.2  \; {\rm GeV}^{2}, $
$ \alpha_{\rm s}(0) = 0.35, $ Ref. [3].
(b) $ m_{\rm b} = 4.898 \; {\rm GeV} $,
$ \sigma = 0.18 \; {\rm GeV}^{2}, $
Shirkov-Solovtsov $ \alpha_{\rm s}( Q^{2} ) $.
(c) $ m_{\rm b} = 4.7605 \; {\rm GeV} $,
$ \sigma = 0.18 \; {\rm GeV}^{2}, $
Dokshitzer $ et \; al. \; \alpha_{\rm s}( Q^{2} ) $.}
\begin{tabular}{cccccc}
 States &  & exp. & (a) & (b) & (c) \\
  &  & (GeV) & (GeV) & (GeV) & (GeV) \\
\hline \hline
$ 1 \, {^{1} {\rm S}_{0}} $ &  &  &
9.374 & 9.374 & 9.375 \\
$ 1 \, {^{3} {\rm S}_{1}} $ &
$ \Upsilon(1S) $ & $ 9.46037 \pm 0.00021 $ &
9.460 & 9.460 & 9.460 \\
$ 2 \, {^{1} {\rm S}_{0}} $ &  &  &
9.975 & 9.988 & 9.983 \\
$ 2 \, {^{3} {\rm S}_{1}} $ &
$ \Upsilon(2S) $ & $ 10.02330 \pm 0.00031 $ &
10.010 & 10.023 & 10.017 \\
$ 3 \, {^{1} {\rm S}_{0}} $ &  &  &
10.322 & 10.342 & 10.328 \\
$ 3 \, {^{3} {\rm S}_{1}} $ &
$ \Upsilon(3S) $ & $ 10.3553 \pm 0.0005 $ &
10.348 & 10.368 & 10.352 \\
$ 4 \, {^{1} {\rm S}_{0}} $ &  &  &
10.598 & 10.618 & 10.391 \\
$ 4 \, {^{3} {\rm S}_{1}} $ &
$ \Upsilon(4S) $ & $ 10.5800 \pm 0.0035 $ &
10.620 & 10.639 & 10.612 \\
$ 5 \, {^{1} {\rm S}_{0}} $ &  &  &
10.837 & 10.854 & 10.816 \\
$ 5 \, {^{3} {\rm S}_{1}} $ &
$ \Upsilon(10860) $ & $ 10.865 \pm 0.008 $ &
10.857 & 10.872 & 10.834 \\
$ 6 \, {^{1} {\rm S}_{0}} $ &  &  &
11.060 & 11.070 & 11.026 \\
$ 6 \, {^{3} {\rm S}_{1}} $ &
$ \Upsilon(11020) $ & $ 11.019 \pm 0.008 $ &
11.079 & 11.089 & 11.044 \\
\hline
$ 1 \, {^{1} {\rm P}_{1}} $ &
  &  & 9.908 & 9.918 & 9.914 \\
$
\begin{array}{c}
 1 \, {^{3} {\rm P}_{2}} \\
 1 \, {^{3} {\rm P}_{1}} \\
 1 \, {^{3} {\rm P}_{0}}
\end{array}
$ & $
\begin{array}{c}
 \chi_{b2}(1P) \\
 \chi_{b1}(1P) \\
 \chi_{b0}(1P)
\end{array}
$ & $ \left.
\begin{array}{c}
 9.9132 \pm 0.0006 \\
 9.8919 \pm 0.0007 \\
 9.8598 \pm 0.0013
\end{array}
\right\} 9.900 $ & 9.908 & 9.920 & 9.917 \\
$ 2 \, {^{1} {\rm P}_{1}} $ &
  &  & 10.260 & 10.279 & 10.269 \\
$
\begin{array}{c}
 2 \, {^{3} {\rm P}_{2}} \\
 2 \, {^{3} {\rm P}_{1}} \\
 2 \, {^{3} {\rm P}_{0}}
\end{array}
$ & $
\begin{array}{c}
 \chi_{b2}(2P) \\
 \chi_{b1}(2P) \\
 \chi_{b0}(2P)
\end{array}
$ & $ \left.
\begin{array}{c}
 10.2685 \pm 0.0004 \\
 10.2552 \pm 0.0005 \\
 10.2321 \pm 0.0006
\end{array}
\right\} 10.260 $ & 10.260 & 10.280 & 10.271
\end{tabular}
\label{tablbb}
\end{table}
\begin{table}
\centering
\caption{$ c \bar{c} $.
$ n_{f} = 4, \; \Lambda = 0.2 $ GeV.
(a) $ m_{\rm c} = 1.394 \; {\rm GeV} $,
$ \sigma = 0.2  \; {\rm GeV}^{2}, $
$ \alpha_{\rm s}(0) = 0.35, $ Ref. [3].
(b) $ m_{\rm c} = 1.545 \; {\rm GeV} $,
$ \sigma = 0.18 \; {\rm GeV}^{2}, $
Shirkov-Solovtsov $ \alpha_{\rm s}( Q^{2} ) $.
(c) $ m_{\rm c} = 1.383 \; {\rm GeV} $,
$ \sigma = 0.18 \; {\rm GeV}^{2}, $
Dokshitzer $ et \; al. \; \alpha_{\rm s}( Q^{2} ) $.}
\begin{tabular}{cccccc}
 States &  & exp. & (a) & (b) & (c) \\
  &  & (MeV) & (MeV) & (MeV) & (MeV) \\
\hline \hline
$ 1 \, {^{1} {\rm S}_{0}} $ &
$ \eta_{c}(1S) $ & $ 2979.8 \pm 2.1 $ &
2982 & 2977 & 2982 \\
$ 1 \, {^{3} {\rm S}_{1}} $ &
$ J/\psi(1S) $ & $ 3096.88 \pm 0.04 $ &
3097 & 3097 & 3097 \\
$ 1 \, \Delta $SS &  & 117 &
115 & 119 & 116 \\
$ 2 \, {^{1} {\rm S}_{0}} $ &
$ \eta_{c}(2S) $ & $ 3594 \pm 5 $ &
3575 & 3606 & 3573 \\
$ 2 \, {^{3} {\rm S}_{1}} $ &
$ \psi(2S) $ & $ 3686.00 \pm 0.09 $ &
3642 & 3670 & 3636 \\
$ 2 \, \Delta $SS &  & 92 &
67 & 64 & 63 \\
$ 3 \, {^{1} {\rm S}_{0}} $ &
  &  &
3974 & 4005 & 3950 \\
$ 3 \, {^{3} {\rm S}_{1}} $ &
$ \psi(4040) $ & $ 4040 \pm 10 $ &
4025 & 4054 & 3998 \\
$ 4 \, {^{1} {\rm S}_{0}} $ &
  &  &
4298 & 4323 & 4252 \\
$ 4 \, {^{3} {\rm S}_{1}} $ &
$ \psi(4415) $ & $ 4415 \pm 6 $ &
4341 & 4364 & 4291 \\
\hline
$ 1 \, {^{1} {\rm P}_{1}} $ &
  &  & 3529 & 3556 & 3528 \\
$
\begin{array}{c}
 1 \, {^{3} {\rm P}_{2}} \\
 1 \, {^{3} {\rm P}_{1}} \\
 1 \, {^{3} {\rm P}_{0}}
\end{array}
$ & $
\begin{array}{c}
 \chi_{c2}(1P) \\
 \chi_{c1}(1P) \\
 \chi_{c0}(1P)
\end{array}
$ & $ \left.
\begin{array}{c}
 3556.17 \pm 0.13 \\
 3510.53 \pm 0.12 \\
 3415.1  \pm 1.0
\end{array}
\right\} 3525 $ & 3530 & 3561 & 3531 \\
$ 2 \, {^{1} {\rm P}_{1}} $ &
&  & 3925 & 3954 & 3904 \\
$ 2 \, {^{3} {\rm P}_{~}} $ &
&  & 3927 & 3958 & 3906 \\
\hline
$ 1 \, {^{1} {\rm D}_{2}} $ &
  &  & 3813 & 3853 & 3811 \\
$
\begin{array}{c}
 1 \, {^{3} {\rm D}_{3}} \\
 1 \, {^{3} {\rm D}_{2}} \\
 1 \, {^{3} {\rm D}_{1}}
\end{array}
$ & $
\begin{array}{c}
     \\
 \psi(3836) \\
 \psi(3770)
\end{array}
$ & $ \left.
\begin{array}{c}
     \\
 ~\,3836 \pm 13 \\
 3769.9 \pm 2.5
\end{array}
\right\} $ & 3813 & 3854 & 3811 \\
$ 2 \, {^{1} {\rm D}_{2}} $ &
  &  & 4149 & 4183 & 4121 \\
$
\begin{array}{c}
 2 \, {^{3} {\rm D}_{3}} \\
 2 \, {^{3} {\rm D}_{2}} \\
 2 \, {^{3} {\rm D}_{1}}
\end{array}
$ & $
\begin{array}{c}
     \\
     \\
 \psi(4160)
\end{array}
$ & $ \left.
\begin{array}{c}
     \\
     \\
 4159 \pm 20
\end{array}
\right\} $ & 4149 & 4184 & 4121
\end{tabular}
\label{tablcc}
\end{table}
\begin{table}
\centering
\caption{$ q \bar{q}. \; m_{\rm u,d} = 0.01 \; {\rm GeV},
\; n_{f} = 4 $.
(a) $ \alpha_{\rm s}(0) = 0.35, $ Ref. [3],
$ \sigma = 0.2 \; {\rm GeV}^{2} $,
$ \Lambda = 0.2 $ GeV.
(b) Shirkov-Solovtsov $ \alpha_{\rm s}( Q^{2} ) $,
$ \sigma = 0.18 \; {\rm GeV}^{2} $,
$ \Lambda = 0.2 $ GeV.
(c) Dokshitzer $ et \; al. \; \alpha_{\rm s}( Q^{2} ) $,
$ \sigma = 0.18 \; {\rm GeV}^{2} $,
$ \Lambda = 0.2 $ GeV.}
\begin{tabular}{cccccc}
 States &  & exp. & (a) & (b) & (c) \\
  &  & (MeV) & (MeV) & (MeV) & (MeV) \\
\hline \hline
$ 1 \, {^{1} {\rm S}_{0}} $ &
$
\left\{
\begin{array}{c}
 \pi^{0} \\
 \pi^{\pm}
\end{array}
\right.
$ & $
\left.
\begin{array}{c}
 134.9764  \pm 0.0006 \\
 139.56995 \pm 0.00035
\end{array}
\right\}
$ &
479 & - & 575 \\
$ 1 \, {^{3} {\rm S}_{1}} $ &
$ \rho(770) $ & $ 768.5 \pm 0.6 $ &
846 & 423 & 904 \\
$ 1 \, \Delta $SS &  & 630 &
367 & - & 329 \\
$ 2 \, {^{1} {\rm S}_{0}} $ &
$ \pi(1300) $ & $ ~\, 1300 \pm 100 $ &
1326 & 952 & 1338 \\
$ 2 \, {^{3} {\rm S}_{1}} $ &
$ \rho(1450) $ & $ 1465 \pm 25 $ &
1461 & 1128 & 1459 \\
$ 2 \, \Delta $SS &  & 165 &
135 & 176 & 121 \\
$ 3 \, {^{1} {\rm S}_{0}} $ &
$ \pi(1800) $ & $ 1795 \pm 10 $ &
1815 & 1485 & 1793 \\
$ 3 \, {^{3} {\rm S}_{1}} $ &
$ \rho(2150) $ & $ 2149 \pm 17 $ &
1916 & 1600 & 1889 \\
$ 3 \, \Delta $SS &  & 354 &
101 & 115 & 96 \\
\hline
$
\begin{array}{c}
 1 \, {^{1} {\rm P}_{1}} \\
 1 \, {^{3} {\rm P}_{2}} \\
 1 \, {^{3} {\rm P}_{1}} \\
 1 \, {^{3} {\rm P}_{0}}
\end{array}
$
&
$
\begin{array}{c}
 b_{1}(1235) \\
 a_{2}(1320) \\
 a_{1}(1260) \\
 a_{0}(1450)
\end{array}
$
&
$
\begin{array}{c}
 1231 \pm 10 \\
 \left.
 \begin{array}{c}
  1318.1 \pm 0.7 \\
  ~\, 1230 \pm 40 \\
  ~\, 1450 \pm 40
 \end{array}
 \right\} 1303
\end{array}
$
&
1333 & 1045 & 1365 \\
\hline
$
\begin{array}{c}
 1 \, {^{1} {\rm D}_{2}} \\
 1 \, {^{3} {\rm D}_{3}} \\
 1 \, {^{3} {\rm D}_{2}} \\
 1 \, {^{3} {\rm D}_{1}}
\end{array}
$
&
$
\begin{array}{c}
 \pi_{2}(1670) \\
 \rho_{3}(1690) \\
      ~   \\
 \rho(1700)
\end{array}
$
&
$
\begin{array}{c}
 1670 \pm 20 \\
 \left.
 \begin{array}{c}
  1691.1 \pm 5 \\
     ~   \\
  1700 \pm 20
 \end{array}
 \right\}
\end{array}
$
&
1701 & 1444 & 1715 \\
\hline
$
\begin{array}{c}
 1 \, {^{1} {\rm F}_{3}} \\
 1 \, {^{3} {\rm F}_{4}} \\
 1 \, {^{3} {\rm F}_{3}} \\
 1 \, {^{3} {\rm F}_{2}}
\end{array}
$
&
$
\begin{array}{c}
    ~  \\
 a_{4}(2040) \\
    X(2000) \\
    ~
\end{array}
$
&
$
\begin{array}{c}
    ~ \\
 \left.
 \begin{array}{c}
  2037 \pm 26 \\
    ~  \\
    ~
 \end{array}
 \right\}
\end{array}
$
&
1990 & 1743 & 1985 \\
\hline
$
\begin{array}{c}
 1 \, {^{1} {\rm G}_{4}} \\
 1 \, {^{3} {\rm G}_{5}} \\
 1 \, {^{3} {\rm G}_{4}} \\
 1 \, {^{3} {\rm G}_{3}}
\end{array}
$
&
$
\begin{array}{c}
      ~ \\
 \rho_{5}(2350) \\
      ~ \\
 \rho_{3}(2250)
\end{array}
$
&
$
\begin{array}{c}
  ~ \\
 \left.
 \begin{array}{c}
  2330 \pm 35 \\
   ~ \\
   ~
 \end{array}
 \right\}
\end{array}
$
&
2238 & 1994 & 2214 \\
\hline
$
\begin{array}{c}
 1 \, {^{1} {\rm H}_{5}} \\
 1 \, {^{3} {\rm H}_{6}} \\
 1 \, {^{3} {\rm H}_{5}} \\
 1 \, {^{3} {\rm H}_{4}}
\end{array}
$
&
$
\begin{array}{c}
     ~ \\
 a_{6}(2450) \\
     ~ \\
     ~
\end{array}
$
&
$
\begin{array}{c}
     ~ \\
 \left.
 \begin{array}{c}
  2450 \pm 130 \\
     ~  \\
     ~
 \end{array}
 \right\}
\end{array}
$
&
2460 & 2215 & 2416
\end{tabular}
\label{tabluu}
\end{table}
\begin{table}
\caption{$ q \bar{q}. \; n_{f} = 4 $.
Shirkov-Solovtsov $ \alpha_{\rm s}( Q^{2} ). \;
\sigma = 0.18 \; {\rm GeV}^{2}. \;
\Lambda = 0.2 \; {\rm GeV} $.
(d) $ m_{\rm u,d} = 0.22 \; {\rm GeV} $.
(e) $ m_{\rm u,d} = 0.30 \; {\rm GeV} $.
(f) $ m_{\rm u,d}^{2} = 0.11 \,
k - 0.025 \, k^{2} + 0.265 \, k^{4} $.}
\begin{tabular}{cccccc}
 States & (MeV) & exp. & (d) & (e) & (f) \\
\hline \hline
$ 1 \, {^{1} {\rm S}_{0}} $ &
$
\left\{
\begin{array}{c}
 \pi^{0} \\
 \pi^{\pm}
\end{array}
\right.
$ & $
\left.
\begin{array}{c}
 134.9764  \pm 0.0006 \\
 139.56995 \pm 0.00035
\end{array}
\right\}
$ & 26 & 473 & 124 \\
$ 1 \, {^{3} {\rm S}_{1}} $ &
$ \rho(770) $ & $ 768.5 \pm 0.6 $ &
725 & 868 & 737 \\
$ \Delta $ SS &  & 630 & 699 & 394 & 613 \\
$ 2 \, {^{1} {\rm S}_{0}} $ &
$ \pi(1300) $ & $ 1300 \pm 100 $ &
1190 & 1326 & 1401 \\
$ 2 \, {^{3} {\rm S}_{1}} $ &
$ \rho(1450) $ & $ 1465 \pm 25 $ &
1344 & 1468 & 1508 \\
$ \Delta $ SS &  & 165 & 154 & 142 & 107 \\
$ 3 \, {^{1} {\rm S}_{0}} $ &
$ \pi(1800) $ & $ 1795 \pm 10 $ &
1688 & 1806 & 1993 \\
$ 3 \, {^{3} {\rm S}_{1}} $ &
$ \rho(2150) $ & $ 2149 \pm 17 $ &
1788 & 1900 & 2063 \\
$ \Delta $ SS &  & 354 & 100 & 94 & 70 \\
\hline
$
\begin{array}{c}
 1 \, {^{1} {\rm P}_{1}} \\
 1 \, {^{3} {\rm P}_{2}} \\
 1 \, {^{3} {\rm P}_{1}} \\
 1 \, {^{3} {\rm P}_{0}}
\end{array}
$
&
$
\begin{array}{c}
 b_{1}(1235) \\
 a_{2}(1320) \\
 a_{1}(1260) \\
 a_{0}(1450)
\end{array}
$
&
$
\begin{array}{c}
 1231 \pm 10 \\
 \left.
 \begin{array}{c}
  1318.1 \pm 0.7 \\
  1230 \pm 40 \\
  1450 \pm 40
 \end{array}
 \right\} 1303
\end{array}
$
& 1243 & 1364 & 1319 \\
\hline
$
\begin{array}{c}
 1 \, {^{1} {\rm D}_{2}} \\
 1 \, {^{3} {\rm D}_{3}} \\
 1 \, {^{3} {\rm D}_{2}} \\
 1 \, {^{3} {\rm D}_{1}}
\end{array}
$
&
$
\begin{array}{c}
 \pi_{2}(1670) \\
 \rho_{3}(1690) \\
      ~   \\
 \rho(1700)
\end{array}
$
&
$
\begin{array}{c}
 1670 \pm 20 \\
 \left.
 \begin{array}{c}
  1691.1 \pm 5 \\
     ~   \\
  1700 \pm 20
 \end{array}
 \right\}
\end{array}
$
& 1603 & 1715 & 1741 \\
\hline
$
\begin{array}{c}
 1 \, {^{1} {\rm F}_{3}} \\
 1 \, {^{3} {\rm F}_{4}} \\
 1 \, {^{3} {\rm F}_{3}} \\
 1 \, {^{3} {\rm F}_{2}}
\end{array}
$
&
$
\begin{array}{c}
    ~  \\
 a_{4}(2040) \\
    X(2000) \\
    ~
\end{array}
$
&
$
\begin{array}{c}
    ~ \\
 \left.
 \begin{array}{c}
  2037 \pm 26 \\
    ~  \\
    ~
 \end{array}
 \right\}
\end{array}
$
& 1881 & 1979 & 2043 \\
\hline
$
\begin{array}{c}
 1 \, {^{1} {\rm G}_{4}} \\
 1 \, {^{3} {\rm G}_{5}} \\
 1 \, {^{3} {\rm G}_{4}} \\
 1 \, {^{3} {\rm G}_{3}}
\end{array}
$
&
$
\begin{array}{c}
      ~ \\
 \rho_{5}(2350) \\
      ~ \\
 \rho_{3}(2250)
\end{array}
$
&
$
\begin{array}{c}
  ~ \\
 \left.
 \begin{array}{c}
  2330 \pm 35 \\
   ~ \\
   ~
 \end{array}
 \right\}
\end{array}
$
& 2118 & 2209 & 2319 \\
\hline
$
\begin{array}{c}
 1 \, {^{1} {\rm H}_{5}} \\
 1 \, {^{3} {\rm H}_{6}} \\
 1 \, {^{3} {\rm H}_{5}} \\
 1 \, {^{3} {\rm H}_{4}}
\end{array}
$
&
$
\begin{array}{c}
     ~ \\
 a_{6}(2450) \\
     ~ \\
     ~
\end{array}
$
&
$
\begin{array}{c}
     ~ \\
 \left.
 \begin{array}{c}
  2450 \pm 130 \\
     ~  \\
     ~
 \end{array}
 \right\}
\end{array}
$
& 2329 & 2415 & 2569
\end{tabular}
\label{shrkuu}
\end{table}

\end{document}